\documentclass[aps,pra,reprint,amsmath,amssymb,superscriptaddress]{revtex4-1}
\usepackage{amsmath}
\usepackage{times}
\usepackage[english]{babel}
\usepackage[caption = false]{subfig}
\usepackage[colorlinks=true,linkcolor=blue,urlcolor=blue,citecolor=blue]{hyperref}
\usepackage{color}
\usepackage{amscd}
\usepackage{amsfonts}
\usepackage{amssymb}
\usepackage{graphicx, epstopdf}
\usepackage{tabularx}
\usepackage{braket}
\usepackage{bm}

\newcommand{\teqref}[1]{Eq.~\eqref{#1}}

\newcommand{\tcite}[1]{~\cite{#1}}
\newcommand{\tref}[1]{~\ref{#1}}
\newcommand{\ov}{``}

\newlength\mytemplen
\newsavebox\mytempbox

\makeatletter
\newcommand\mybluebox{%
    \@ifnextchar[
       {\@mybluebox}%
       {\@mybluebox[0pt]}}

\def\@mybluebox[#1]{%
    \@ifnextchar[
       {\@@mybluebox[#1]}%
       {\@@mybluebox[#1][0pt]}}

\def\@@mybluebox[#1][#2]#3{
    \sbox\mytempbox{#3}%
    \mytemplen\ht\mytempbox
    \advance\mytemplen #1\relax
    \ht\mytempbox\mytemplen
    \mytemplen\dp\mytempbox
    \advance\mytemplen #2\relax
    \dp\mytempbox\mytemplen
    \colorbox{myblue}{\hspace{1em}\usebox{\mytempbox}\hspace{1em}}}

\begin{document}

\title{Quantum dynamics in the interacting Fibonacci chain}

\author{Cecilia Chiaracane}
\email{chiaracc@tcd.ie}
\affiliation{School of Physics, Trinity College Dublin, Dublin 2, Ireland}
\author{Francesca Pietracaprina}
\email{pietracf@tcd.ie}
\affiliation{School of Physics, Trinity College Dublin, Dublin 2, Ireland}
\author{Archak Purkayastha}
\email{archak.p@tcd.ie}
\affiliation{School of Physics, Trinity College Dublin, Dublin 2, Ireland}
\author{John Goold}
\email{gooldj@tcd.ie}
\affiliation{School of Physics, Trinity College Dublin, Dublin 2, Ireland}

\begin{abstract}
Quantum dynamics on quasiperiodic geometries has recently gathered significant attention in ultracold atom experiments where nontrivial localized phases have been observed. One such quasiperiodic model is the so-called Fibonacci model. In this tight-binding model, noninteracting particles are subject to on-site energies generated by a Fibonacci sequence. This is known to induce critical states, with a continuously varying dynamical exponent, leading to anomalous transport. In this work, we investigate whether anomalous diffusion present in the noninteracting system survives in the presence of interactions and establish connections to a possible transition towards a localized phase. We investigate the dynamics of the interacting Fibonacci model by studying real-time spread of density-density correlations at infinite temperature using the dynamical typicality approach. We also corroborate our findings by calculating the participation entropy in configuration space and investigating the expectation value of local observables in the diagonal ensemble.

\end{abstract}
\date{\today}
\maketitle

\section{Introduction}
Anderson localisation is the phenomenon where electrons undergo quantum coherent scattering with random impurities and give rise to a metal-insulator transition. In three spatial dimensions, a critical energy depending on the disorder strength, the so-called mobility edge, separates localized from extended states associated with diffusion~\tcite{Semeghini2015}. In one dimension, instead, all the eigenstates get spatially localized when even an infinitesimal amount of uncorrelated disorder is introduced, and the wires switch from ballistic conductors to insulators~\tcite{gangof4}. Yet if the random disorder is replaced by a quasiperiodic potential, incommensurate with the underlying periodicity of the lattice, a wider variety of behaviors arises even in one dimension. The paradigmatic example is Aubry-Andr\'e-Harper (AAH) model, where the cosine potential, modified by an irrational factor in its argument, induces a transition from a completely delocalized phase to a completely localized phase as the strength of the quasiperiodic potential is increased~\tcite{aubry1980analyticity, hiramoto}. 

At the critical point of the noninteracting AAH model both the spectrum and the eigenfunctions show a fractal structure, and transport becomes subdiffusive~\tcite{purkayastha2018,vkv2017,Sutradhar2019, purkayastha2019}.  Recently, there have been several theoretical works investigating the effect of many-body interactions on the AAH model \cite{yoo2020,Cookmeyer2020,Znidaric_interacting_AAH_2018,BarLev2017,Naldesi2016,Mastropietro2015,settino1,Iyer2013,Tezuka2012,Zhong_1995}. Moreover, the control over the Hamiltonian and the initial conditions in ultracold atom setups has given the platform to realize quasiperiodic models and probe their nontrivial transport properties~\tcite{exp2} from the perspective of dynamics. In these experiments, by tuning the relative depths of the optical lattices trapping the atoms, it is possible to investigate both the noninteracting limit of the models as well as the effect of many-body interactions on the dynamics. For example, the single particle localization in the AAH model is known to give rise to many-body localization (MBL)~\tcite{expmbloc, expmbloc3, expmbloc4,luschen2017slow} in presence of interactions.

Another well known example of a quasiperiodic system is the Fibonacci model, characterized by a potential generated by the Fibonacci substitution rule. Though this model is topologically related to the AAH model \cite{Kraus2012,Goblot2020}, its transport properties are known to be extremely different. For example there is no delocalization-localization transition in the Fibonacci model. Instead, the Fibonacci potential induces critical behavior of all eigenstates at every potential strength \cite{Kohmoto1987, hiramoto1988, mace2016, vkv2017, Zhong_1995,jagannathan2020fibonacci}. The transport exponents show a smooth crossover from ballistic to subdiffusive with increase in the potential strength~\tcite{hiramoto1988, Fibotransport, varma2019}. A natural question, then, is what happens to the transport behavior of the Fibonacci model in the presence of interactions?

Different answers to this question have been proposed in the literature, ranging from a transition towards MBL~\tcite{mace2019}, to metal-insulator transitions at low energies \cite{PhysRevLett.83.3908,PhysRevB.65.014201}, to persistence of the anomalous diffusion~\tcite{settino}. Motivated by the lack of experimental results at this moment, in this paper, we focus on characterizing the transport of the interacting Fibonacci model with a further approach, exploiting dynamical quantum typicality~\tcite{typ1, typ2}. We study the real-time broadening of the expectation values of local number operators, starting from a nonequilibrium typical state, that can act as representative of the equilibrium ensemble. Via dynamical quantum typicality, the quantities are directly related to the spread of density-density correlations and thereby to classification of transport via the Green-Kubo formula in the isolated system~\tcite{typ3,typ4,stein2017}. This approach allows us to access larger system sizes and much longer time scales than otherwise possible. We find strong evidence of subdiffusive transport at large enough potential strengths, which precedes a crossover to a possible MBL phase. We further corroborate the crossover to this phase by means of a study of participation entropy and calculations in the diagonal ensemble.
 
 This article is structured as follows. In Sec.\tref{sec:nonint}, we present the noninteracting Fibonacci model and display its nontrivial transport properties by reproducing the known results of the spreading of an initially localized wave packet. In Sec.\tref{sec:int}, we add many-body interactions to the Hamiltonian, and review the previous works in the literature about the interacting Fibonacci model. Dynamical quantum typicality is introduced in Sec.\tref{sec:dqt}, where we also show explicitly how to exploit it in order to compute the infinite temperature density correlations we use to classify transport and present our main results. In Sec.\tref{sec:ed} and\tref{sec:de}, we complete the observations from the study of the dynamics with a further investigation respectively on the participation entropy of the system and the expectation values of both the local occupation and imbalance in diagonal ensemble. We conclude and summarize in Sec.\tref{sec:concl}.

\section{Noninteracting Fibonacci model}
\label{sec:nonint}
The Fibonacci model is a one-dimensional system of noninteracting fermions, described by the Hamiltonian
\begin{equation}
\label{eq:hni}
\hat{H}_{\text{NI}}=\sum\limits_{l=1}^{N-1}  t_h \  (\hat{a}^{\dagger}_{l}\hat{a}_{l+1}+{\rm h.c})+ \sum\limits_{l=1}^{N} u_l\hat{a}^{\dagger}_l\hat{a_l},
\end{equation}
where $\hat{a}_l$ is the annihilation operator of a fermion on site $l$, $t_h$ is the tunneling constant, and $u_l$ is the on-site energy of site $l$. The on-site potential is binary $u_l= (u_A, u_B)$, and the chain of values on the $N$ sites is obtained by repeatedly applying the Fibonacci substitution rule, given by 
\begin{align}
        u_A & \rightarrow u_A u_B  \\
        u_B & \rightarrow u_A.
\end{align}
Alternatively, the sequence of $N$ on-site energies can be built connecting together two smaller sequences. Starting from two initial chains $C_0 = [ u_B], \ C_1 =  [u_A]$, one gets $C_2 = C_1C_0 = [u_A, u_B]$. Increasingly longer chains are generated by concatenation of the segments from the two previous generations  $C_n = [C_{n-1}, C_{n-2}]$. Consequently, the length of every chain $C_n$ belongs to the Fibonacci sequence $F_n \in \{1, 1,2,3,5, 8, \dots \}$. Unlike a periodic sequence, generated by smaller parts of it and presenting the same rate of $u_A$ to $u_B$ even in the indefinitely extended limit, the Fibonacci chain exhibits a $u_A/u_B$ ratio equal to $F_n/F_{n-1}$ at the $n$th generation, which goes to $1/\tau$ for $n \rightarrow \infty$,  with $\tau = (1+\sqrt{5})/2$ the golden ratio~\tcite{goodson2017}. Experiments or simulations can be strongly limited in system size and usually involve chains of generic length $N$ that do not belong to the Fibonacci chain. To treat the model with the Fibonacci potential within such small system sizes, we adopt the averaging procedure used in Refs.~\cite{varma2019,mace2016}.  We consider an \ov infinite'' sequence with $N_{\infty} \gg N$ and cut out finite samples of length $N$. It is possible to prove that there exist $N+1$ nonequivalent samples, among which one (two) is reflection symmetric around the center of the chain for $N$ even ($N$ odd) and each of the remaining configurations has a symmetric partner, with same eigenvalues and eigenfunctions~\tcite{goodson2017}. Therefore, after discarding the reflection symmetric examples and the symmetric partners of the samples already considered, $N/2$ [or $(N-1)/2$] distinct samples are available to average over. This averaging procedure also restores effective translational invariance in the thermodynamic limit. The quasiperiodicity of the potential gives rise to a multifractal spectrum at every $u_A$ and $u_B$~\tcite{Kohmoto1987, mace2016}. Therefore, we assume without loss of generality to control a single parameter $u_A = - u_B = u$ in units of $t_h$. It has been shown that the multifractality of the spectrum induces anomalous behavior in the transport properties of the noninteracting system \tcite{hiramoto1988}. 

\begin{figure}
\centering
\subfloat[]{\includegraphics[width=0.95\columnwidth]{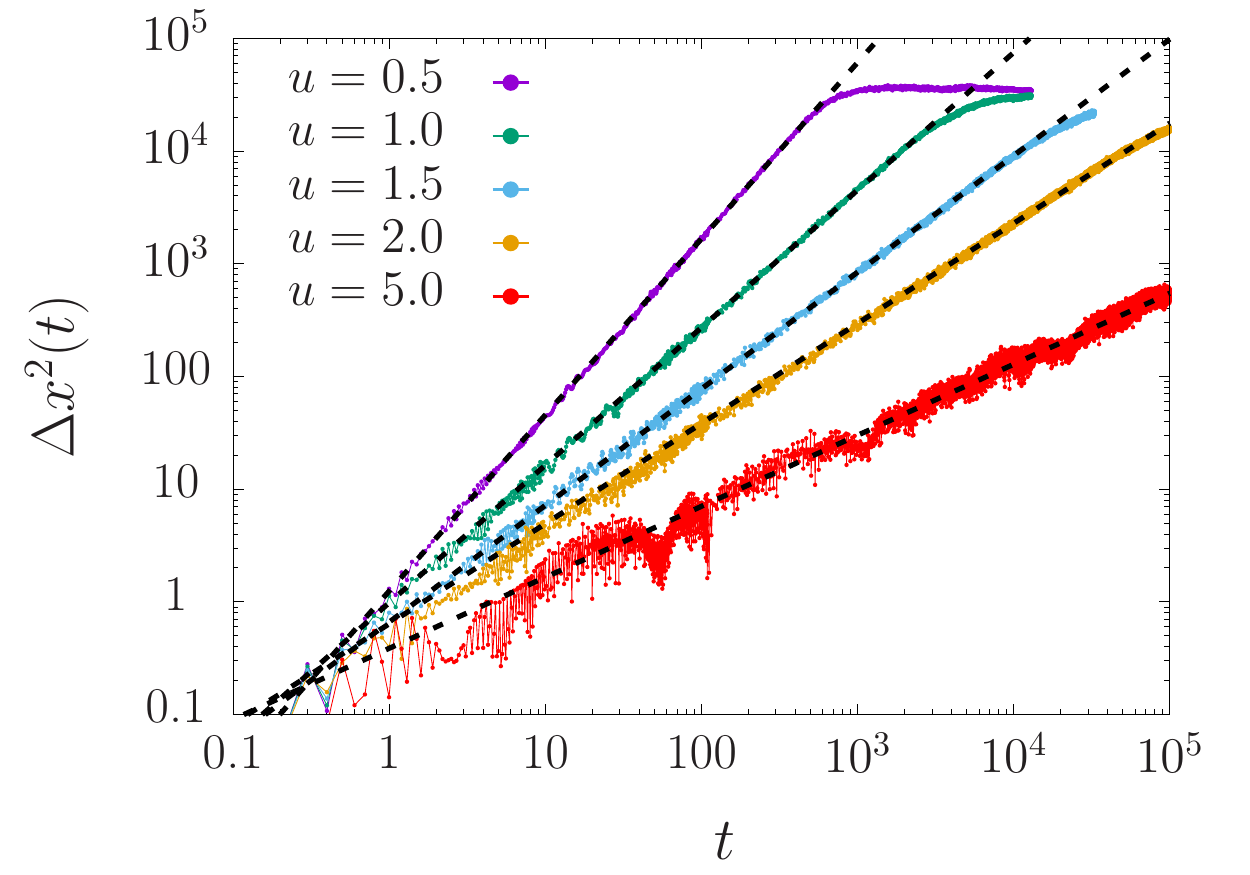} \label{fig:dx2ni}} \\
\subfloat[]{\includegraphics[width=0.95\columnwidth]{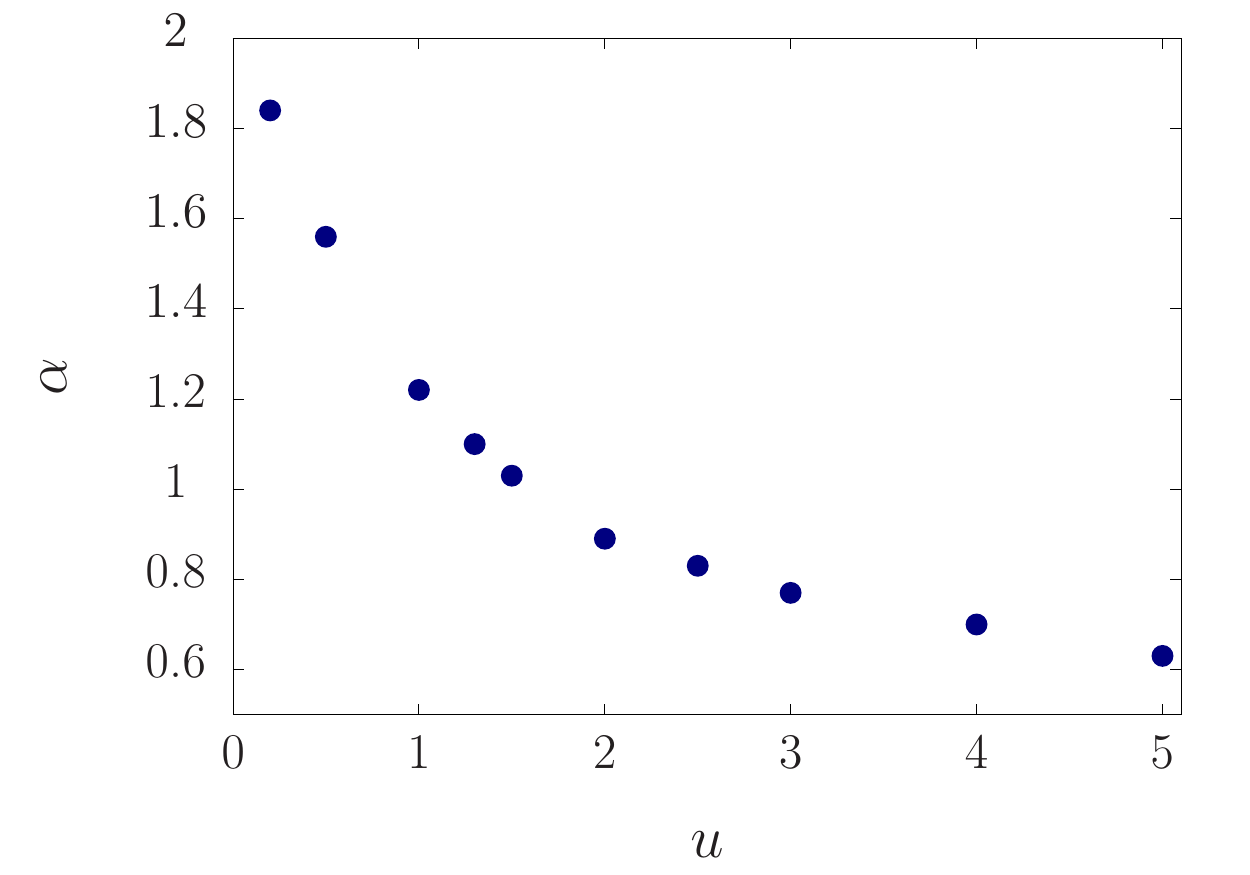} \label{fig:expni}}
\caption{(a) Mean squared displacement $\Delta x^2$ of a state initially localized at the middle of a noninteracting Fibonacci chain of $N=1001$ sites is computed in time for different example values of the potential strength $u$. The fits shown by the dashed lines in the log-log plot follows \teqref{eq:dfit}. We notice the curves saturating at low $u$ as the size of the system is reached, where the fast dynamics makes the finite size of the system visible at shorter times. (b) The extracted exponent $\alpha$ varies continuously with $u$, indicating anomalous diffusion.}
\end{figure}

For completeness, we start here by reproducing a known result on the transport properties of the noninteracting Fibonacci model. We initialize the isolated system in open boundary condition (OBC) with a fermion on site $N/2$. Thus the chosen initial state at $t=0$ is $\ket{\Psi(0)} = \sum_l \Psi_l(0) \ \hat{a}^{\dagger}_l \ket{0}$, with $\Psi_l(0) = \delta_{l N/2}$ the components over the site basis and $\ket{0}$ the vacuum state. We trace the time evolution of the components $\Psi_l(t)$, governed by the Schr\"{o}dinger equation, and the mean squared displacement of the wave function
\begin{equation}
\label{eq:dx2}
\Delta x^2(t) = \sum_l \bigl[ (l- N/2)^2 \  \lvert \Psi_l(t) \rvert^2 \bigr].
\end{equation}
The asymptotic time dependence of this quantity can be written as
\begin{equation}
\label{eq:dfit}
\Delta x^2(t) \sim t^{\alpha},
\end{equation}
where $\alpha = 0$ implies localization, $\alpha = 2$ denotes ballistic transport, $\alpha = 1$ implies diffusion, $1<\alpha<2$ implies superdiffusive transport, and $0<\alpha<1$ implies subdiffusive transport. In Fig.\tref{fig:dx2ni} we show the $\Delta x^2(t)$ as a function of time for a chain of $N= 1001$, averaged over the $(N-1)/2$ nonequivalent configurations, for different values of the potential strength $u$. Because of the finite system size, at long enough times, $\Delta x^2(t)$ saturates. This saturation happens at time scales where the initially localized wave packet has spread over the entire system. For small $u$, this time scale goes as $\sim N/2$. As $u$ increases, the transport slows down, and correspondingly the saturation happens at longer times. The dynamical exponent $\alpha$ corresponds to the thermodynamic limit behavior, and needs to be obtained from the long time behavior of the system before this saturation happens.  In Fig.\tref{fig:expni} we show $\alpha$, extracted from the fits of the $\Delta x^2(t)$ curves, as a function of $u$. As expected, $\alpha$ tends to $2$ and transport becomes ballistic for $u \rightarrow 0$. It then decreases continuously towards $0$ for increasing $u$, with two regimes: superdiffusive ($\alpha > \/1$ for $u \lesssim 1.5$) and subdiffusive ($\alpha < \/ 1$ for $u \gtrsim 1.5$). The above characterization of transport in the noninteracting Fibonacci model is well known. The main results of this work concern the Fibonacci model in the presence of interactions, which we describe below.

\section{the Interacting Fibonacci model}
\label{sec:int}
The interacting Fibonacci model is realized by adding to \teqref{eq:hni} a nearest neighbor density-density term
\begin{equation}
\label{eq:Hint}
\hat{H}_{\text{I}}  = \hat{H}_{\text{NI}} +  2\Delta \sum\limits_{l=1}^{N-1} \hat{n}_{l+1}\hat{n}_l,
\end{equation}
where $\hat{n}_l = \hat{a}_l^{\dagger}\hat{a}_l$ is the fermionic number operator on site $l$ and $\Delta$ is the strength of the many-body interaction. 
 
The interacting version of the Fibonacci model has recently started to receive attention with respect to how the many-body term affects the transport properties of the system. In the recent work by Varma and \v{Z}nidari\v{c}, the dynamics of polarized domain walls and the boundary-driven Lindblad equation steady states reveal diffusion at small interaction strengths\tcite{varma2019}. The spectral analysis in Ref.\tcite{mace2019} provides, instead, evidence for a localization transition at finite potential strength, that would constitute a genuine many-body effect, since the noninteracting model does not exhibit a localized phase. Finally, a nonequilibrium  Green’s functions approach in Ref.\tcite{settino} suggests in the Fermi-Hubbard realization of the model a slow subdiffusive behavior at high potential strength, determined by the nontrivial spectral properties of the model.

Here we aim to investigate the survival of anomalous diffusion in the interacting model. In the presence of interactions, the classification of transport based on the spread of a localized wave packet is no longer possible. Instead, we classify transport by the spread of an inhomogeneity on the infinite temperature state of the system. Via dynamical typicality, as we show below, this is exactly analogous to the spread of a localized wave packet, and reduces to the same in the absence of interactions. 

\section{Dynamical quantum typicality}
\label{sec:dqt}
In many-body problems, the numerical simulation of the Schr\"odinger equation is technically challenging due the exponential growth of the Hilbert space dimension $D$ with the number of degrees of freedom of the system. Popular techniques, such as the time-dependent density matrix renormalization group\tcite{paeckel2019}, can push the simulations to large system sizes $N \sim 200$, but are limited to short times due to the increase of entanglement, and cannot generally reach the time scales required to study equilibrium properties. However, it is possible to exploit the concept of dynamical quantum typicality (DQT) to circumvent part of these difficulties. The approximation tells us that it is possible to infer the dynamics of the system from a single pure state $\ket{\psi}$ drawn at random on an arbitrary basis $\{ \ket{\phi_k} \}_{k=1}^{D}$, that can be considered as ``typical" representative of the statistical ensemble~\tcite{typ1, typ2}, as we explain below.

We write explicitly the typical state as
\begin{equation}
    \ket{\psi} = \hat{R} \sum_{k=1}^D c_k \ket{\phi_k}, \ \ \ \ \ \ \ c_k = a_k + ib_k,
\end{equation}
with $\hat{R}$ an arbitrary linear operator and  $a_k$ and $b_k$ mutually independent random variables from Gaussian distributions with zero mean and variance 1/2. It can be shown from the properties of the coefficients that the averaged expectation value of an arbitrary Hermitian operator $\hat{O}$ in the typical state is equivalent to the expectation value taken with respect to a density matrix $\hat{\rho}$, as follows
\begin{align}
\overline{O} &= \overline{\braket{\psi|\hat{O}| \psi}} = Tr[ \hat{\rho} \hat{O}],
\end{align}
where the overline indicates the average over the probability distribution, and the density matrix is defined as
\begin{equation}
\label{eq:densm}
\hat{\rho} = \hat{R} \hat{R}^{\dagger}.
\end{equation}
Since the density matrix is positive semi-definite, it can always be written in the above form. Thus, any mixed state $\hat{\rho}$ can be represented in terms of an ensemble of typical pure states. When we further look at the sample to sample fluctuations in taking the  average over the distribution, assuming $\hat{O}$ is Hermitian, we get
\begin{equation}
\sigma^2_O = \overline{\bigl(\braket{\psi|\hat{O}| \psi} \bigr)^2 } - (\overline{O})^2 =Tr[(\hat{\rho} \hat{O})^2].
\end{equation}
The expression can be bounded from above by
\begin{equation}
\sigma^2_O \le Tr[ \hat{\rho}^2\hat{O}^2] \le \|\hat{O}\|^2 Tr[\hat{\rho}^2],
\end{equation}
with $\|\hat{O}\|^2 =  \bigl(\max{\lambda^0_n}\bigr)^2$ considering the eigenvalues $\{\lambda_n^o\}$ of the operator and $Tr[\hat{\rho}^2]$ the purity of the state $\hat{\rho}$. The above result can also be generalized to cases where $\hat{O}$ is not Hermitian, by breaking $\hat{O}$ into Hermitian and anti-Hermitian parts. For a highly mixed state in a high dimensional Hilbert space $Tr[\hat{\rho}^2] \ll 1$. In such cases, the sample to sample fluctuations in doing the ensemble average also become small, so that, for a large enough system size, a small number of  realizations is enough to calculate expectation values of operators.   
In particular, in the infinite temperature limit, the state is completely mixed, and essentially one typical state realization can be used as representative of the whole ensemble:
\begin{equation}
\label{eq:thetyp}
\hat{\rho} = \frac{1\!\!1}{D},  \ \ \ \ \ \ \ \ket{\psi} = \frac{1}{\sqrt{D}} \sum\limits_{k=1}^D c_k \ket{\phi_k},
\end{equation}
where $D=2^N$ for the Hamiltonian $\hat{H}_{\text{I}}$.

The formulation described above does not depend on any specific property of the operator $\hat{O}$. It can also be a combination of operators in the Heisenberg picture so that both the dynamics as well as two-time correlations can be obtained. Dynamical typicality can be used to connect two-time density correlations in the infinite temperature state to dynamics following an initially localized quench over the infinite temperature state,  as we explain in the following section. 

\subsection{Density correlations at infinite temperature and spread of a localized quench}
\label{sec:denscorr}
Two-time density correlations are intimately connected with transport properties of an isolated system in the thermodynamic limit \cite{PhysRevB.99.144422,PhysRevResearch.2.013130}. The infinite temperature correlation function reads
\begin{align}
C_{l q} (t) &=  \braket{\hat{n}_l(t)\hat{n}_q} - \braket{\hat{n}_l}\braket{\hat{n}_q} \nonumber \\ &= \frac{Tr[\hat{n}_l(t)\hat{n}_q]}{2^N} - \frac{Tr[\hat{n}_l]}{2^N}\frac{Tr[\hat{n}_q]}{2^N}.
\end{align}
Let us choose $q = N/2$, and define $C_l (t) = C_{l N/2}(t)$. We can then further simplify by using $\hat{n}^2_{N/2} = \hat{n}_{N/2}$ and $Tr[\hat{n}_p] = 2^{N-1}$, as shown in the following
\begin{align}
C_l(t) &= \frac{Tr[\hat{n}_l(t)\hat{n}_{N/2}]}{2^N} - \frac{Tr[\hat{n}_l]}{2^N}\frac{Tr[\hat{n}_{N/2}]}{2^N} \nonumber \\ &= \frac{Tr[\hat{n}_{N/2}\hat{n}_l(t)\hat{n}_{N/2}]}{2^N} - \frac{1}{4} \nonumber \\
&=\overline{\braket{\psi_{N/2}| \hat{n}_l(t)|\psi_{N/2}}}- \frac{1}{4}
\end{align}
In the last equality, we have exploited typicality with
\begin{equation}
\ket{\psi_{N/2}} = \hat{n}_{N/2} \ket{\psi}, \ \ \ \ \ \ \  \ket{\psi} = \frac{1}{2^{N/2}} \sum_{k=1}^D c_k \ket{\phi_k}
\end{equation}
where, as seen previously in \teqref{eq:thetyp}, $\ket{\psi}$ is the typical state associated to the thermal state at infinite temperature. Since the dimension of the Hilbert space grows exponentially, at large enough sizes $N \gtrsim 20$ sample to sample fluctuations are negligible and it can be considered only one typical state. When we normalize the typical state to $\ket{\tilde{\psi}_{N/2}} \approx \sqrt{2}\ket{\psi_{N/2}}$ (shown in Appendix~\tref{app:halfill}, which follows Refs.\tcite{PhysRevB.99.144422, PhysRevResearch.2.013130, stein2017}), we finally obtain
\begin{equation}
\label{eq:vediapp}
C_l(t) \approx C^{typ}_l(t) = \frac{1}{2} \Bigl( n_l(t) - \frac{1}{2} \Bigr),
\end{equation}
with $n_l(t) = \braket{\tilde{\psi}_{N/2}| \hat{n}_l(t)|\tilde{\psi}_{N/2}}$. Thus the density-density correlation $C_l(t)$ is given by the dynamics of the expectation value of $\hat{n}_l$ after a quench induced by the normalized projection of a typical state onto the subspace where the site $N/2$ is occupied. The subtraction of 1/2 within the parentheses amounts to subtracting the background initial occupation of sites away from the middle of the chain $l \neq N/2$. In general, as computed explicitly in Appendix~\tref{app:halfill}, we have
\begin{equation}
\label{eq:Ctyp}
C_l(t) \approx C^{typ}_l(t) = \frac{1}{2} \Bigl( n_l(t) -  n_l(0) \Bigr), \ \ \ l \neq N/2
\end{equation}
where, again, the expectation values are evaluated  with respect to $\ket{\tilde{\psi}_{N/2}}$.

\begin{figure*}
\centering
\subfloat[]{\includegraphics[width=0.24\textwidth]{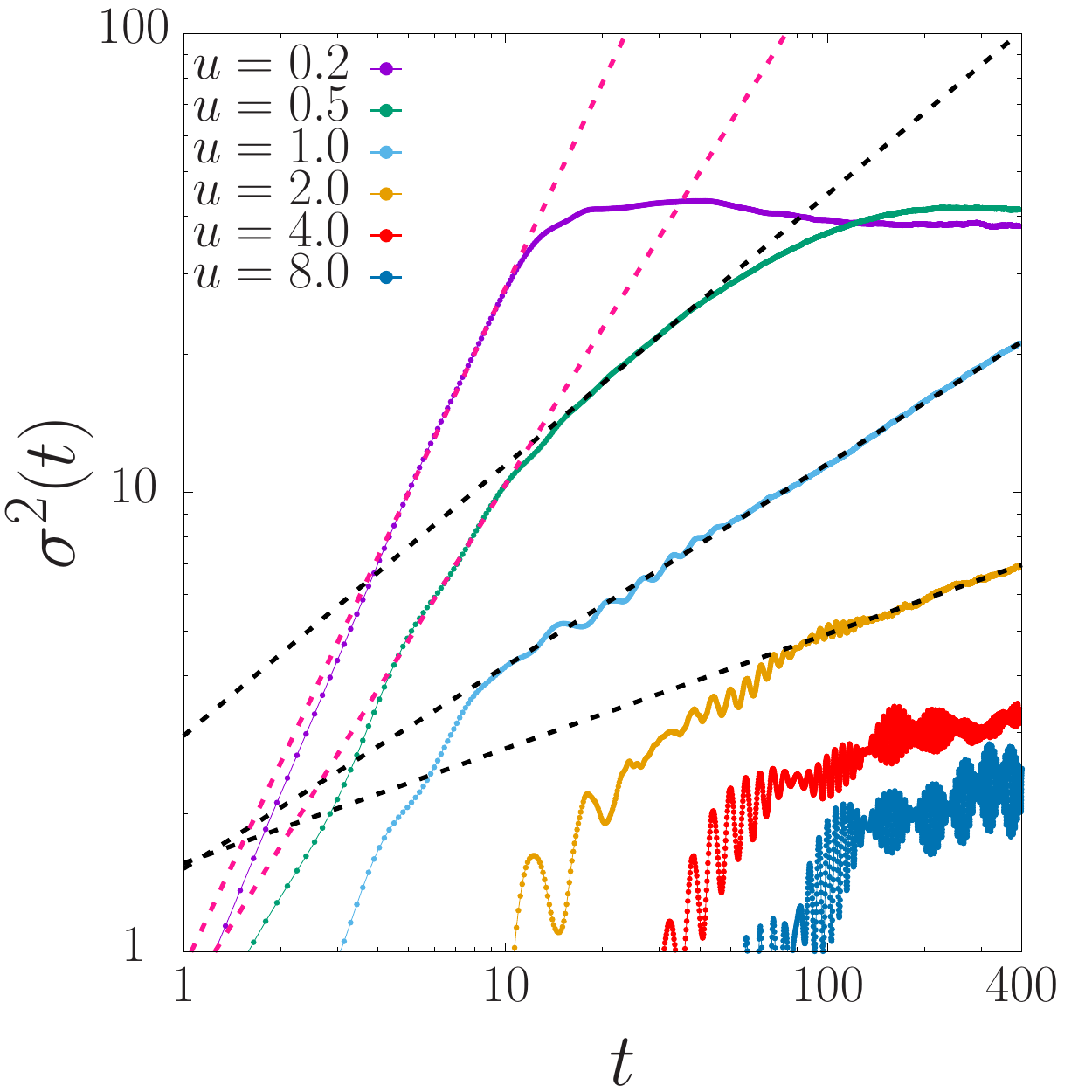} \label{fig:sigmai}}
\subfloat[]{\includegraphics[width=0.24\textwidth]{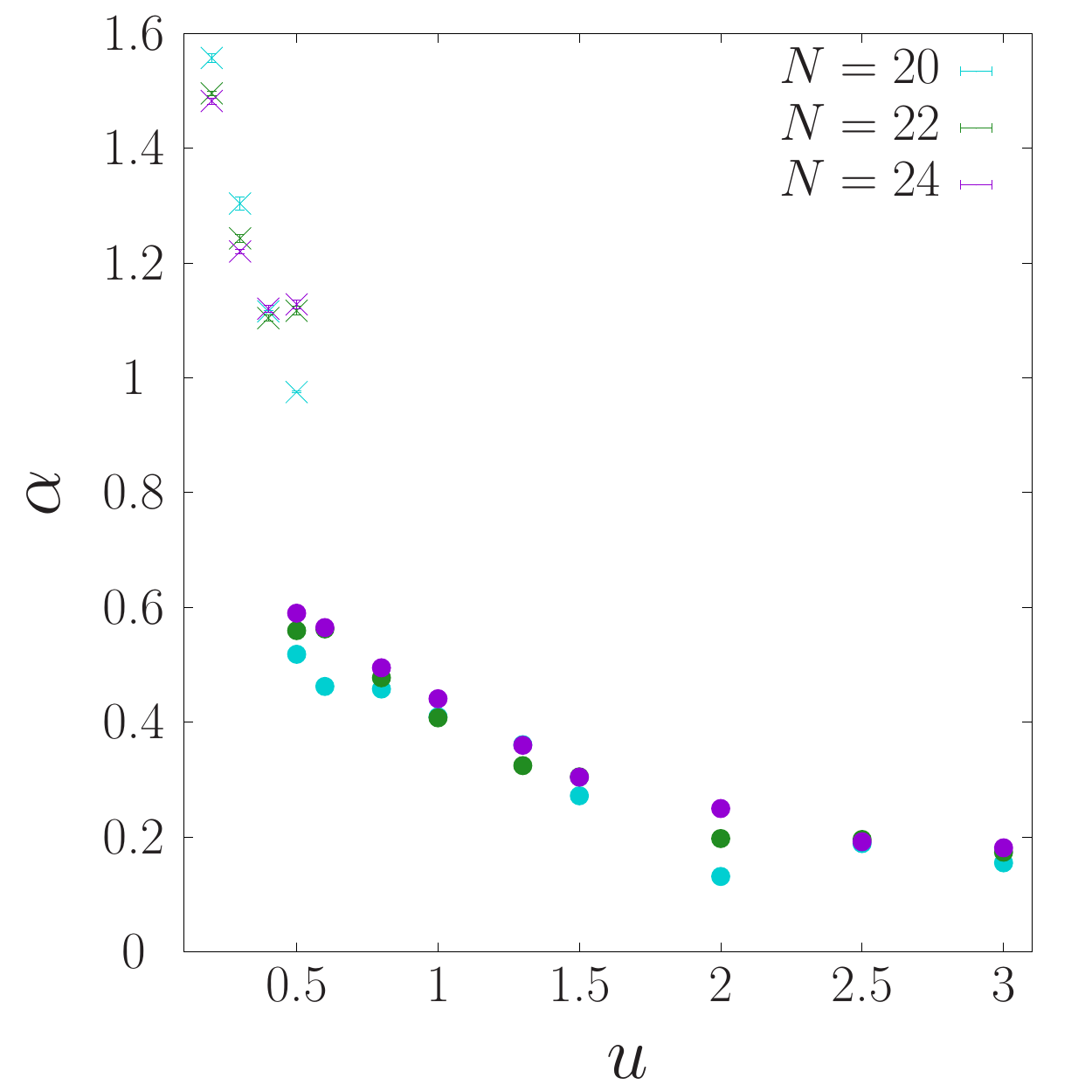}
\label{fig:alphai}}
\subfloat[]{\includegraphics[width=0.24\textwidth]{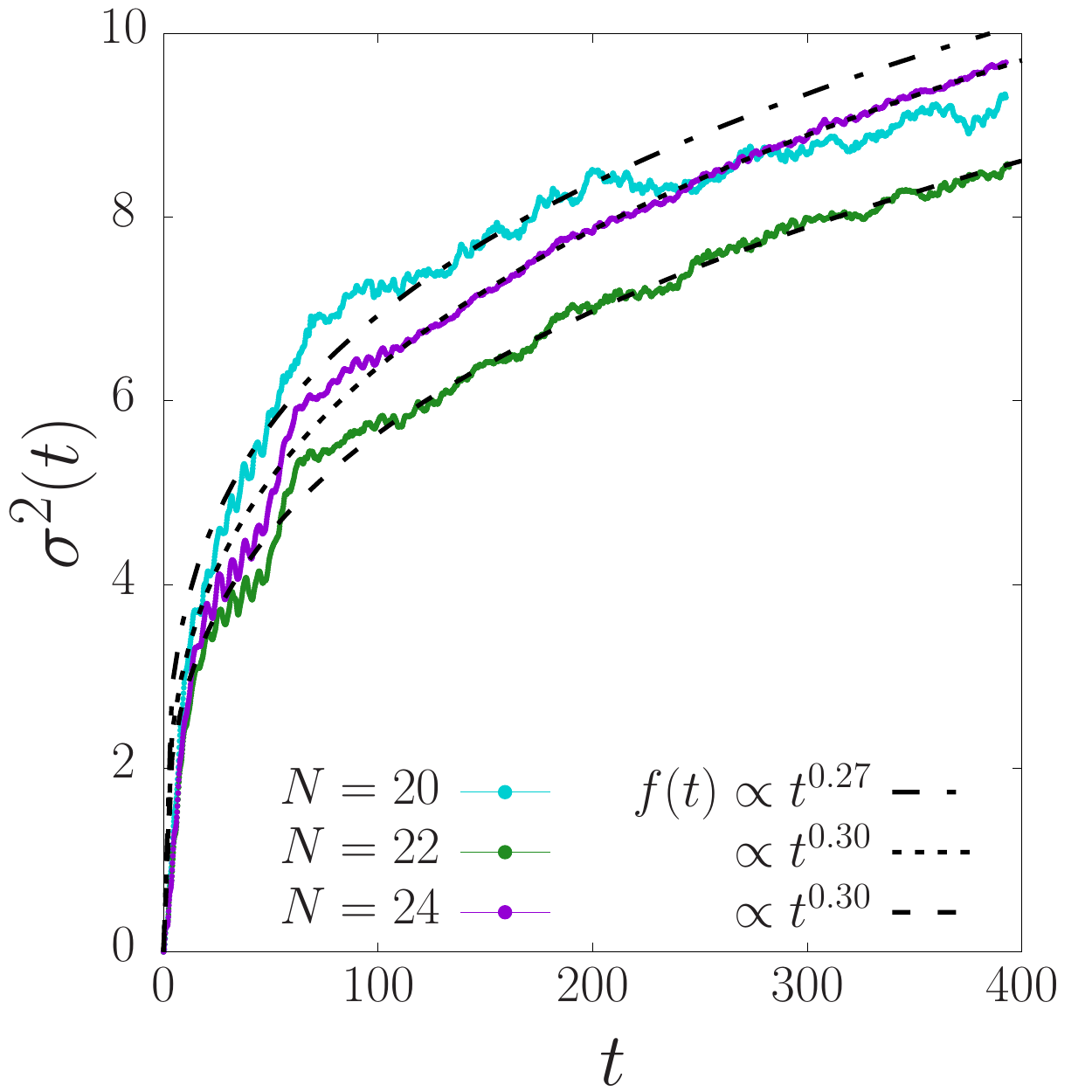} \label{fig:exemplei}}
\subfloat[]{\includegraphics[width=0.24\textwidth]{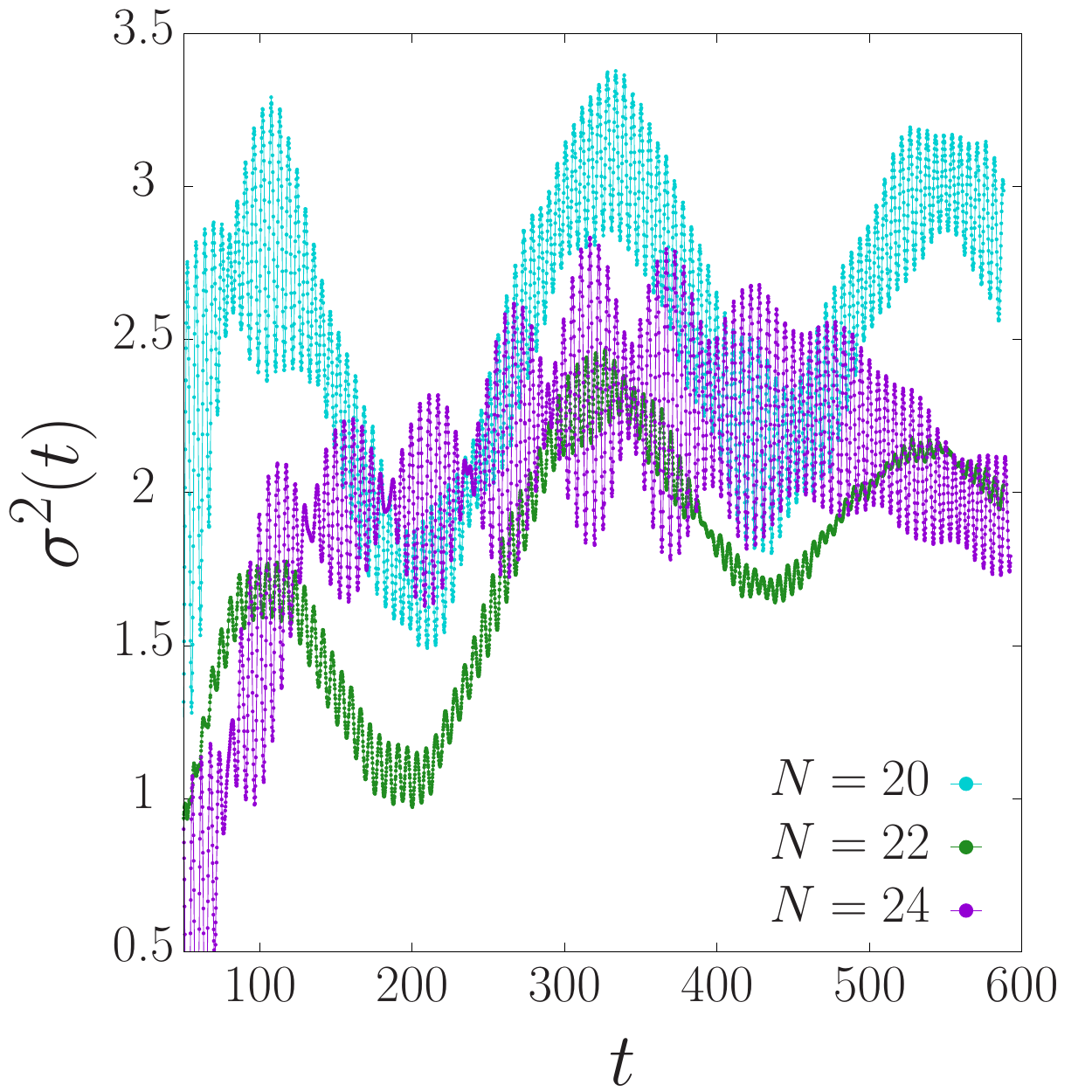} \label{fig:u8}}
\caption{(a) Log-log plot of $\sigma^2(t)$ vs time $t$ at different Fibonacci potential strengths $u$, for a chain of $N=24$ spins and $\Delta=0.5$. The data computed from the Krylov time evolution are shown with continuous lines. We show the corresponding fits of the form of \teqref{eq:dfit2} in dotted lines; the short-time fits are shown in pink, while the long-time ones are in black. We notice $\sigma^2(t)$ growing faster than $t$ at low $u$, and slowing down as $u$ increases. (b) Exponent $\alpha$ extracted from the $\sigma^2(t)$ as a function of the potential strength $u$. The crosses correspond to the fast dynamics [shown in pink in panel (a)], while the dots are relative to the long-time dynamics [shown in black in panel (a)]. The errors of each data point are smaller than the dot size. (c) $\sigma^2(t)$ is shown at $u=1.5$ for three different system-sizes in linear scale, along with their corresponding power-law fits. (d) The same quantity is displayed for $u=8.0$ at three different system sizes $N=20,22, 24$ for a longer time in linear scale. }
\end{figure*}

In order to classify transport, we define the spatial variance 
\begin{equation}
\label{eq:sigsq}
\sigma^2(t)    = 4\sum\limits_{l=1}^{N} \left(l-\frac{N}{2}\right)^2 \  C_l ^{typ}(t).
\end{equation}
As shown in Appendix~\ref{app:Green-Kubo}, where we reproduce the derivation in Refs.\tcite{typ3,typ4,stein2017}, this quantity is related to the time-dependent diffusion coefficient at infinite temperature limit,
\begin{align}
& \frac{d\sigma^2}{dt}=8\mathcal{D}(t), \\
& \mathcal{D}(t)=\lim_{\beta \rightarrow 0} \lim_{N\rightarrow \infty}\frac{1}{N}\int_{0}^{t} dt^\prime \langle \hat{I}(t^\prime) \hat{I} \rangle_{\beta,\mu}.
\end{align}
For a diffusive system, the diffusion coefficient is constant, $\mathcal{D}(t)=\mathcal{D}$, and hence $\sigma^2(t) = 8 \mathcal{D} t$.
If $\mathcal{D}(t)$ decreases with time, it points to subdiffusive transport. If $\mathcal{D}(t)$ is zero, then there is no transport. If $\mathcal{D}(t)\sim t$, then the transport is ballistic, while if $\mathcal{D}(t)$ diverges with a power less than $1$, then the transport is superdiffusive. Consequently, if, in general,
\begin{align}
\label{eq:dfit2}
\sigma^2(t) \sim t^{\alpha},
\end{align}
then $\alpha=1$ corresponds to diffusive transport, $\alpha=2$ corresponds to ballistic transport, $1<\alpha<2$ corresponds to superdiffusive transport, $0<\alpha<1$ corresponds to subdiffusive transport, and $\alpha=0$ corresponds to lack of transport. This is exactly same as the transport classification in terms of $\Delta x^2(t)$ for the noninteracting system given in \teqref{eq:dfit}. Indeed, as we show in Appendix~\ref{app:relation_with_wavepacket}, for a noninteracting system, $\sigma^2(t)=\Delta x^2(t)$.

Another way to characterize transport is via decay of density autocorrelation with time. The infinite temperature density autocorrelation at site $N/2$ is given by
\begin{align}
&C_{N/2}(t) = \frac{Tr[\hat{n}_{N/2}(t)\hat{n}_{N/2}]}{2^N} - \frac{Tr[\hat{n}_l]}{2^N} \nonumber \\
& \approx C^{typ}_{N/2}(t) = \frac{1}{2} \Bigl( n_{N/2}(t) - \frac{1}{2} \Bigr)
\end{align} 
Thus, via typicality, it corresponds to how the occupation at the middle site approaches its thermal value following the quench. We assume a general power-law decay of  autocorrelation, 
\begin{align}
\label{eq:dfit3}
C_{N/2}(t)\approx C^{typ}_{N/2}(t) \sim t^{-\nu}.
\end{align}
For a ballistic system, $\nu=1$, for a diffusive system $\nu=1/2$, while for a localized system $\nu=0$. Correspondingly, $1<\nu<1/2$ points to superdiffusive transport and  $1/2<\nu<0$ points to subdiffusive transport. For diffusive and ballistic systems, this exponent $\nu$ is related to the exponent $\alpha$ as $\alpha=2\nu$. But, for anomalous transport, these two exponents may not be directly related. In the following, we present results for transport classification of the interacting Fibonacci chain based on calculation of both the exponents.

\subsection{Results}
\label{sec:res}
We numerically study the Hamiltonian, recast through Jordan-Wigner transformations into that of a spin 1/2 XXZ model with external magnetic field, 
\begin{equation}
\label{eq:Hints}
\hat{H}_I =  \sum\limits_{l=1}^{N-1} [ \ t_h (\hat{s}^+_l\hat{s}^-_{l+1} + {\rm h.c} ) + \  2\Delta \  \hat{s}^z_l \hat{s}^z_{l+1}]  + \sum\limits_{l=1}^N u_l \hat{s}^z_l
\end{equation}
with $\hat{s}_l^+$, $\hat{s}_l^-$, $\hat{s}_l^z$ respectively the raising, lowering and $z$ spin operators at site $l$, and $u_l$ the on-site potential following the Fibonacci sequence.  We restrict our calculations to sectors of the total Hilbert space with fixed magnetization, choosing, in particular, the largest one with $N/2$ spins up. Details on how to modify the expressions in the DQT approach  are reported in Appendix~\tref{app:halfill}. We generate a single typical state by taking a normalized state vector $\ket{\tilde{\psi}^s}$ with random coefficients and apply the operator $\hat{n}_{N/2}$.  We time evolve this state using the Krylov subspace method~\tcite{PhysRevLett.51.2238} and calculate the density profile at each time point, from which both $\sigma^2(t)$ and $C_{N/2}(t)$. The dynamical typicality approach, together with the Krylov subspace methods, allows us to do a long time simulation of a maximum system size of $N=24$.  We fix the hopping term $t_h = 1$, and the interaction strength $\Delta = 0.5$, and investigate the nature of transport as a function of the strength of the Fibonacci potential $u$. All the results shown are averaged over the collection of the nonequivalent samples, as described in Sec.\tref{sec:nonint}.

In Fig.\tref{fig:sigmai}, we show $\sigma^2(t)$ as a function of $t$ at significant values of $u$, for $N=24$. As with the noninteracting system, for small $u$, $u \lesssim \Delta $, saturation occurs at time $\sim N/2$ due to the finite size of the system. Power-law fitting of the data before the saturation yields a superdiffusive exponent, $2>\alpha>1$ [see plot for $u=0.2$ in Fig.\tref{fig:sigmai}]. On increasing $u$, the transport slows down, and therefore it takes a much longer time to hit saturation.   For $u \gtrsim \Delta$, we see a clear subdiffusive exponent, $1>\alpha>0$ (plots for $u=1.0,~2.0$ in Fig.~\ref{fig:sigmai}) and 
saturation is not reached within our simulation time scales and system sizes. The crossover from superdiffusive to subdiffusive behavior seems to occur at $u\approx \Delta$, where from our results there does not seem to be a clear power-law behavior before the saturation happens. It is possible, at best, to fit two different power-laws at two different time regimes, between $5\lesssim t \lesssim 10$ with $\alpha \sim 1.1$ and between $10 \lesssim t \lesssim 50$ with $\alpha \sim 0.5$. The exponents obtained from the power-law fits are given in Fig.~\ref{fig:alphai}, which shows the crossover from superdiffusive to subdiffusive transport.   At much higher values of $u$, $u \gg \Delta$, $\sigma^2(t)$ again to quickly saturate to a finite, low value: this points to a lack of spreading of the initially localized quench, $\alpha=0$, thereby pointing at a many-body localized (MBL) regime [$u=4.0, 8.0$ in Fig.~\ref{fig:sigmai}]; this is reminiscent of the results of Ref.~\cite{mace2019} and will be explored in more detail in the following section. As often in literature, it is hard to pinpoint exactly at which values of $u$ the crossover to  a MBL regime happens from dynamical results. 

\begin{figure}
\centering
\includegraphics[width=0.95\columnwidth]{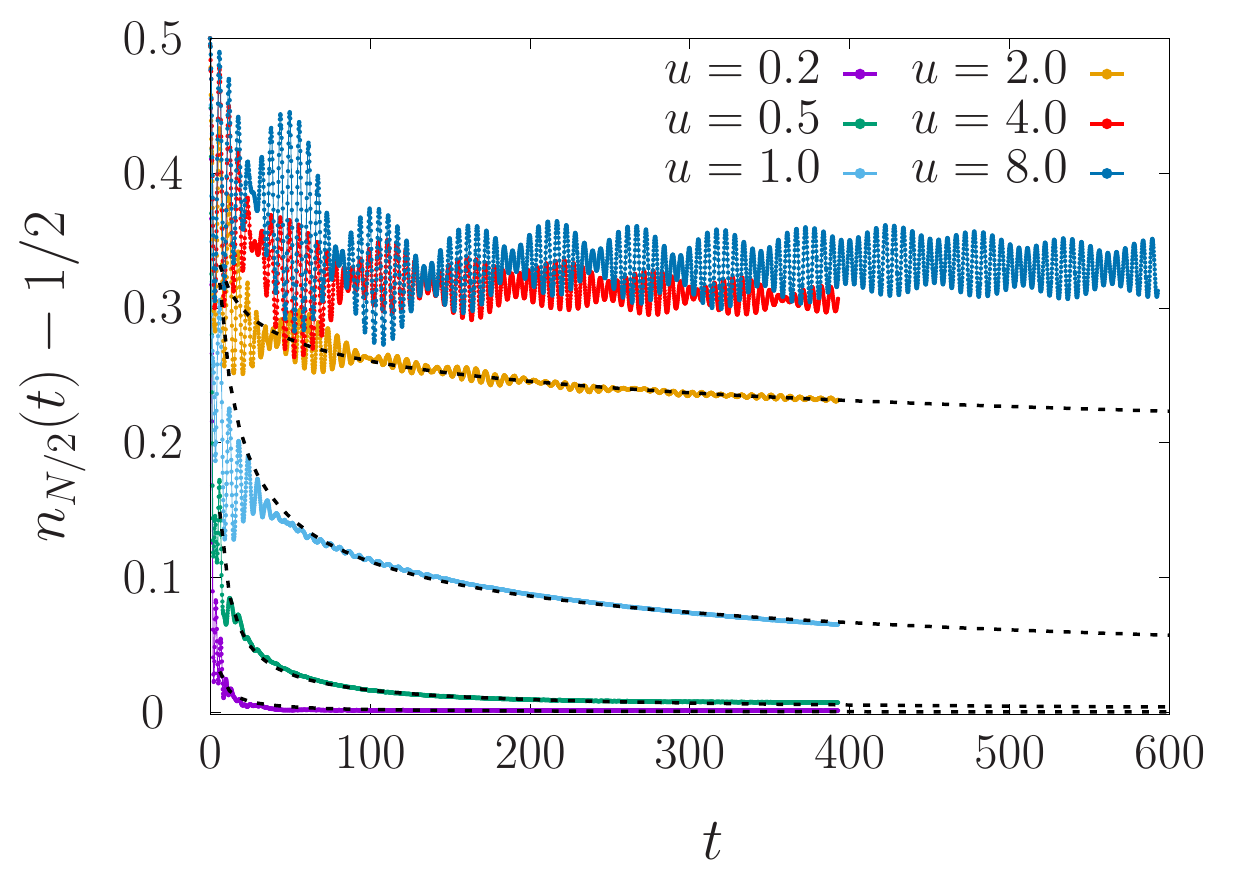}
\caption{Time evolution of $2C_{N/2}(t)=n_{N/2}(t) - \frac{1}{2}$ evaluated on the typical state projected over the subspace where the site at the center of the chain $N/2$ is initially occupied by one particle.  At $u=0.2, 0.5, 1.0, 2.0$, $C_{N/2}(t)$ decays as $t^{-\nu}$, with respectively $\nu = 0.92, 0.78, 0.37, 0.086$. The fits are shown in dashed lines. At high $u$, $C_{N/2}(t)$ does not seem to show any decay up to the longest simulation time.} \label{fig:nhalf}
\end{figure}

To highlight the differences between the subdiffusive and the MBL regime  and to discuss finite-size effects, in Figs.~\ref{fig:exemplei} and \ref{fig:u8}, we show plots of $\sigma^2(t)$ for $u=1.5$ and $u=8.0$, respectively, for different system sizes. In Fig.~\ref{fig:exemplei}, the long time power-law growth of $\sigma^2(t)$ with a subdiffusive exponent is clear for all three system sizes $N=20,22,24$. With increase in system size, the time extent of the power-law growth increases, as expected, and the power-law exponent also converges (to $\alpha=0.3$). However, the different system sizes noticeably do not overlap at any time scale. This is due to the effect of the finite system size coupled with the quasiperiodic potential: results for quasiperiodic systems, even for large system sizes are dependent on the particular choice of system sizes~\tcite{Sutradhar2019,purkayastha2018,vkv2017}, particularly, for the Fibonacci potential, on how different the system-sizes are from Fibonacci numbers. This system-size dependence may be reduced by averaging over samples, but the small number of available samples (equal to the system size $N$) limits the kind of averaging that is possible to perform in our system sizes.
We note that, while this behavior holds for all values of $u$, this does not affect our ability to obtain the power-law exponent and that, nevertheless, all the results for the three different system sizes are of the same order of magnitude.
In Fig.~\ref{fig:u8}, this same size-dependent effect is shown in the localized regime for $u=8.0$.
Here, we highlight the presence of oscillations, while at the same time showing no signs of a power-law growth trend.

Next, we look at $C_{N/2}(t)$ and characterize transport in terms of the exponent $\nu$ [Eq.~(\ref{eq:dfit3})]. The plots of $2C_{N/2}(t)=n_{N/2}(t) - \frac{1}{2}$ are shown in Fig.~\ref{fig:nhalf}. $2C_{N/2}(t)$ shows oscillations on top of a very clear power-law decay for $u<4.0$. For $u\lesssim \Delta$, the power-law exponent is consistent with superdiffusive transport, $1>\nu>0.5$; for $u\gtrsim \Delta$, the power-law exponent is consistent with subdiffusive transport $0.5>\nu>0$. For $u\gg \Delta$, corresponding to $u=4.0,8.0$ in Fig~\ref{fig:nhalf}, we do not see any power-law decay up to the longest time scales that we simulated, thereby suggesting localization. This is consistent with our results from time scaling of $\sigma^2(t)$.

In Ref.~\cite{varma2019}, it was shown that at small $u$ at large enough system sizes, the behavior becomes diffusive. Since our results are limited to much smaller system sizes, we cannot completely rule out that possibility. Nevertheless, in Ref.~\cite{varma2019}, at larger values of $u$ one parameter was reported showing subdiffusive transport. This is completely consistent with the subdiffusive behavior we observe for $u\gtrsim \Delta$.

Our results strongly suggest that anomalous transport survives in the Fibonacci model in the presence of interactions. Moreover, at $u \gg \Delta$, our investigations on dynamics suggests a crossover to MBL. This is very interesting because, in the absence of interactions, there is no localized phase. MBL in the Fibonacci model was previously reported in Ref.~\cite{mace2019}, at a different strength of interaction $\Delta$. For the remainder of the study, we investigate the existence of MBL for our choice of parameters from spectral properties of the Hamiltonian, and calculations in the diagonal ensemble.

\section{Eigenstate properties across different transport regimes}
\label{sec:ed}
\begin{figure}
\centering
\includegraphics[width=0.95\columnwidth]{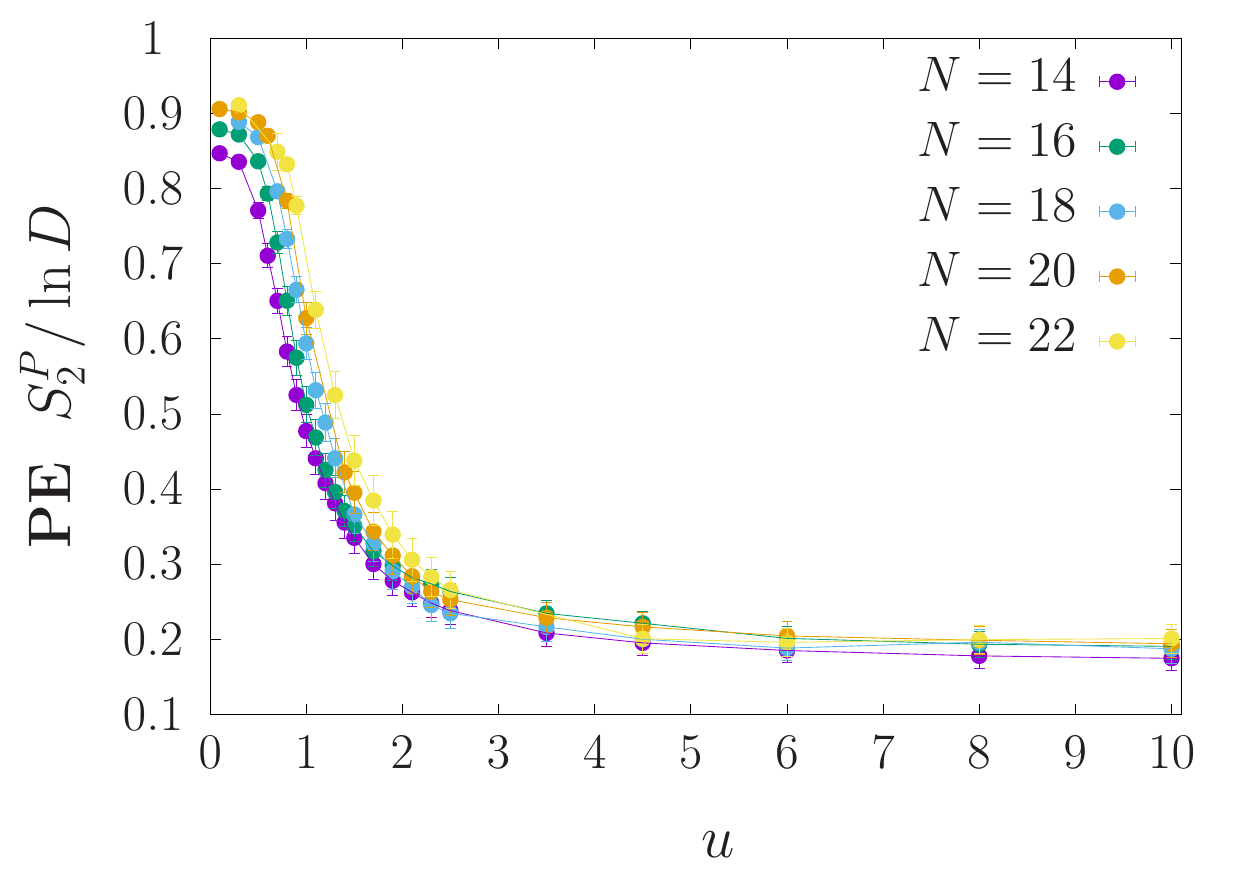}
\caption{Participation entropy $S^P_2/\log{D}$ associated to the central region of the spectrum for different Fibonacci potential strength $u$. The curves are displayed for multiple chain sizes.} \label{fig:PE}
\end{figure}

The spectral properties of the Hamiltonian~\eqref{eq:Hint} have already shown evidence of a many-body localization transition at finite critical potential strength at $\Delta = 1.0$ \tcite{mace2019}. This phase would be introduced uniquely by the interplay of quasidisorder and many-body interactions, since localization is not present in the noninteracting limit of the model. We perform here an analysis similar to \tcite{mace2019}, by computing the R\'enyi-$2$ participation entropy of the Fibonacci chain with $\Delta =0.5$ through exact diagonalization (ED)~\cite{Mac__2019}. These quantities have been used to characterize localization both in single-particle and in many-body interacting systems~\tcite{kramer1993,luitz2014, luitz2015}. 

Let $\ket{\psi_E}$ represent a many-body energy eigenstate. This can be expanded in an arbitrary basis, which we choose to be the configuration-space basis, as $\ket{\psi} = \sum_{k=1}^D d_k \ket{\chi_k}$. The probability $p_k = |d_k|^2$ indicates the \ov participation'' of the element $\ket{\chi_k}$ from the arbitrary basis $\{ \ket{\chi_k} \}_k$ in the state $\ket{\psi_E}$. The second R\'enyi participation entropy (PE) is given by
\begin{equation}
    S^P_2 = -\ln\Bigl(\sum\limits_{k=1}^{D} p_k^2\Bigr).
\end{equation}
If the eigenstate is completely delocalized, $S^P_2/\log(D)\rightarrow 1$. On the other hand, if $S^P_2/\log(D)\rightarrow D_2$, then the eigenstate is fractal with a fractal dimension of $D_2$. For a system showing MBL, the midspectrum energy eigenstates, the region of the Hilbert space that is sampled by the isolated system at infinite temperature,  are expected to be fractal with a low fractal dimension. For systems which are neither completely delocalized nor in MBL, $S^P_2/\log(D)$ for the midspectrum eigenstates may not converge to a constant. The study of $S^P_2$  thereby allows one to capture crossover to MBL.

In Fig.\tref{fig:PE}, we plot $S^P_2/\log(D)$  as a function of the potential strength for different system sizes, $N = 14, 16, 18, 20, 22$. All the points are obtained from an average over $\sim 200$ midspectrum eigenstates, with the exception of the data for $N=22$ that are averaged over $\sim 140$ eigenstates. Finally, the PE are averaged over the nonequivalent realizations of the Fibonacci potential.  The midspectrum eigenvalues and eigenstates are obtained through the shift-invert algorithm\tcite{pietrac2018}. 
At very low $u$, the PE $S^P_2/\log(D)$ is close to $1$, but still shows dependence on the system size. At larger values of $u$, $S^P_2/\log(D)$ decays rapidly with $u$, and eventually shows a collapse for the different system sizes. 
Thus two regimes can be identified, corresponding to the transport (either superdiffusive or subdiffusive) and no transport regimes found in Sec. \ref{sec:res}, the latter reminiscent of the many-body localized phase identified in Ref.~\cite{mace2019}.
More definitive statements about the transition in the thermodynamic limit would require a systematic study of the finite size scaling, which is beyond the purpose of the present work. Instead, in the following, we explore yet another way of characterizing the MBL transition from finite system sizes.

\section{Diagonal ensemble}
\label{sec:de}
In Sec.\tref{sec:res}, we obtained finite-time results for the dynamics of the system at different potential strengths $u$. In this section, we instead focus on the asymptotic results by using full ED and the diagonal ensemble, or infinite time averaged state, to investigate the time infinite limit of $n_{N/2}(t)$ of the isolated system initialized in a nonequilibrium state $\ket{\psi}$, in order to understand if the system reaches absence of transport at high values of $u$ or rather exhibits a region of slow dynamics. Given the use of full ED, we are limited to smaller system sizes than in Sec. \ref{sec:res}.

\begin{figure}
\centering
\subfloat[]{\includegraphics[width=0.95\columnwidth]{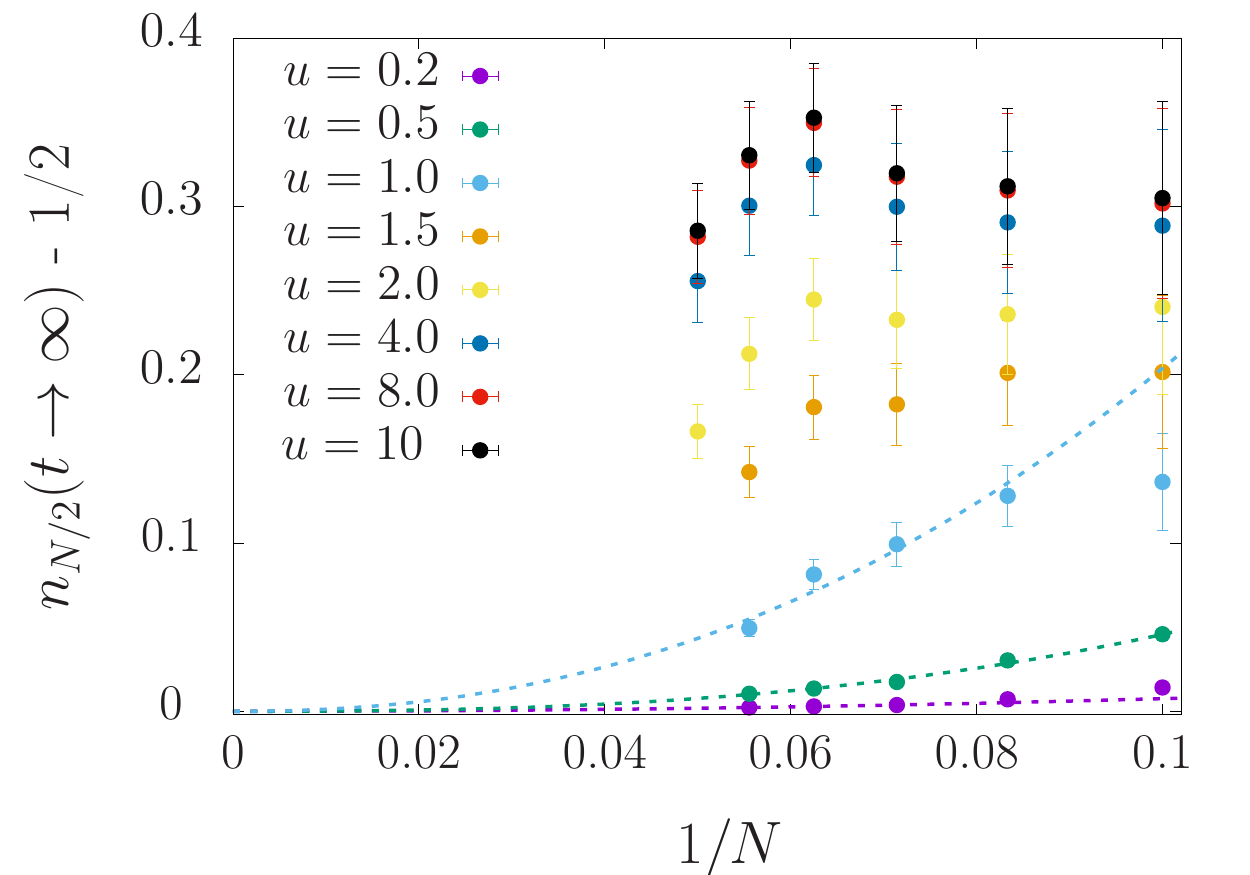} \label{fig:nhalfde}} \\
\subfloat[]{\includegraphics[width=0.95\columnwidth]{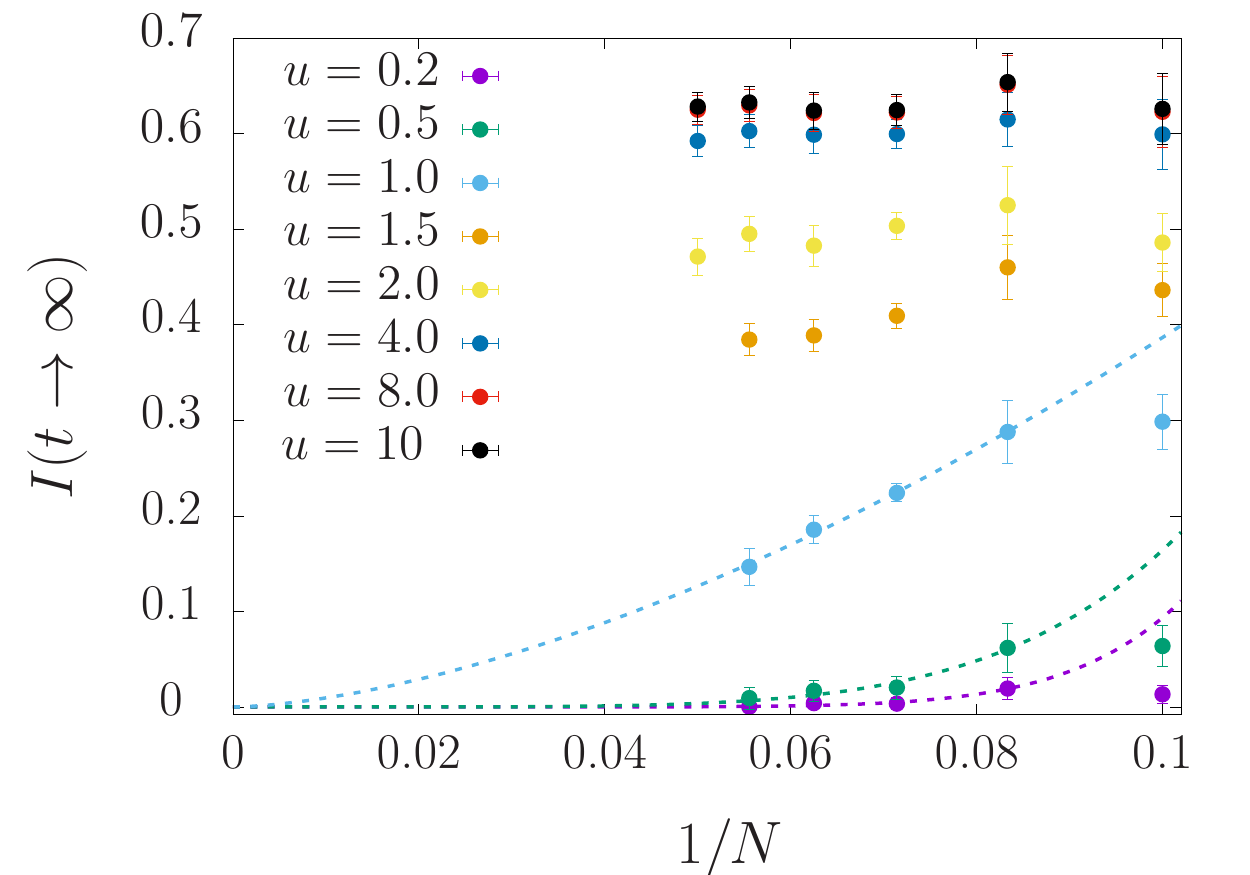} \label{fig:imbde}}
\caption{(a) Expectation value of the occupation at half chain $\hat{n}_{N/2}$ in the diagonal ensemble associated to the initial typical state, which gives the infinite time limit of the operator. The dotted lines indicate extrapolation for $N \rightarrow \infty$ for the first three values of $u$, described by $1/N^{\gamma}$ with $\gamma = 2.81, 2.59, 2.23$ for increasing $u$. (b) Expectation value of the imbalance $\hat{I}$ in the diagonal ensemble for the initial N\'eel state, with the same color code of (a). Again, the dotted lines represent the fits $1/N^{\gamma}$ we use to extrapolate the value of $I(t\rightarrow \infty)$ in the thermodynamic limit, with $\gamma = 8.73, 5.47, 1.61$ for increasing $u$.}
\end{figure}

The time infinite limit of an arbitrary observable $\hat{O}$ reads as
\begin{equation}
O(t \rightarrow \infty) = \lim_{T \rightarrow \infty} \frac{1}{T} \int_{0}^T \braket{\psi(t)| \hat{O} | \psi(t)} dt,
\end{equation}
and, if the spectrum is not degenerate, it can be easily re-written as
\begin{equation}
O(t \rightarrow \infty) = O_{diag} = \sum\limits_k \braket{\phi_k | \hat{O} | \phi_k} | \braket{\phi_k | \psi} |^2,
\end{equation}
where $O_{diag}$ indicates the expectation value of the operator $\hat{O}$ in the diagonal ensemble relative to the initial state $\ket{\psi}$. However, the computation of $O_{diag}$ requires full ED of the Hamiltonian, so our results are limited to the system sizes $N=10, 12, 14, 16, 18$, up to a maximum of $20$ obtained only at $u \ge 2.0$.

We focus on the occupation number at half chain $\hat{n}_{N/2}$, considering the diagonal ensemble for the typical state. The occupation of the initial state is 1 by construction, and it will eventually reach the equilibrium value of $\sim 0.5$ in the case of thermalization. The results are shown in Fig.\tref{fig:nhalfde} for different potential strengths as a function of the inverse of the system size. At low $u \lesssim 1$, the value of the observable at infinite time decreases with $N$ and we are able to extrapolate the infinite-size limit result through a fit of the form $\sim 1/N^\gamma$; the fits are shown with dotted lines in Fig. \ref{fig:nhalfde} and extrapolate to $0.5$,  indicating that the system thermalizes in the thermodynamic limit. At larger $u$, we do not assume a form for the finite size scaling of $n_{N/2}$ and thus we do not extrapolate the infinite-size limit.

We also consider the imbalance\tcite{expmbloc, exp2}, a density correlation function defined as
\begin{equation}
I(t) = \frac{4}{N} \sum\limits_{j=1}^N \braket{ \psi(0) | (\hat{n}_j(T) - 1/2)(\hat{n}(0) - 1/2) | \psi(0) },
\end{equation}
which in the case of initial  N\'eel state $\ket{\psi_{N}}$ can be written as the following operator 
\begin{equation}
\hat{I} = [\hat{n}_e - \hat{n}_o]/N, 
\end{equation}
where $\hat{n}_{e/o} = \sum_{l_{e/o}} \hat{n}_l$ is the number of particles at the even ($e$) or odd ($o$) sites. We compute the initial imbalance $I(t=0) = \braket{\psi_N| \hat{I} | \psi_N}$, and derive its infinite time limit from its expectation values in the diagonal ensemble associated to the N\'eel state. The initial value at $t=0$ is $1$ and will eventually reach the equilibrium value of $0$ if there is thermalization. We show the infinite-time limit of the imbalance in Fig.\tref{fig:imbde} as a function of $1/N$. The results are similar to those from $\hat{n}_{N/2}$ and the random typical state. At low $u$, it is possible to extrapolate the imbalance in the thermodynamic limit, giving $0$. However, at larger potential strength, namely for $u>4$, the data shows a lack of decay with $N$, up to the system sizes we have access to, and supports our results obtained in Sec. \ref{sec:res} pointing to absence of transport in the system and localization.

\section{Conclusions}
\label{sec:concl}

In this work, we have studied the dynamics of density-density correlations at infinite temperature of the Fibonacci model in the presence of nearest neighbor many-body interactions via direct numerical simulation using the DQT approach. The DQT approach, coupled with Krylov subspace method \cite{stein2017,typ3,typ4}, has allowed us to obtain the density correlations for larger system sizes and much longer time scales than otherwise possible. This allowed us to extract the dynamical exponents corresponding to the transport properties of the model. We have further correlated our results with calculations of the participation entropy of the mid-spectrum states, and with exact diagonalization calculations in the diagonal ensemble corresponding to nonequilibrium initial states, at smaller system sizes. We have focused on a fixed interaction strength, $\Delta=0.5 t_h$, and have characterized the transport as a function of the strength of the Fibonacci potential $u$. The following picture emerges from our investigation. For $u\lesssim \Delta$, the transport is relatively fast. We find some evidence of superdiffusion in this regime, although the fast transport and the finite system sizes do not allow us to extract a long time transport exponent. On increasing $u$, the transport slows down, allowing us to extract long time exponents. For $u\gtrsim \Delta$, we find a strong evidence of subdiffusive transport. The crossover from superdiffusive to subdiffusive behavior seems to occur at $u\sim \Delta$, where we are unable to extract a single dynamical exponent. On further increasing $u$, i.e, for $u\gg \Delta$, we find strong evidence that the system crosses over to an MBL phase. The MBL phase is then further corroborated with studies of the participation entropy and the diagonal ensemble, both of which complement the results from study of the dynamics. 

The above picture that emerges from our study contributes towards filling a gap in our present understanding of interacting quasiperiodic systems. Most studies of interacting quasiperiodic systems have focused on the AAH potential, both in theory and in experiment \cite{yoo2020,Cookmeyer2020,Znidaric_interacting_AAH_2018,BarLev2017,Naldesi2016,Mastropietro2015,Iyer2013,Tezuka2012,expmbloc, expmbloc3, expmbloc4,luschen2017slow}.  Though related with the AAH model, the noninteracting Fibonacci model is known to have very different transport properties, which continuously cross over from ballistic to subdiffusive as a function of the strength of the potential \cite{hiramoto1988, Fibotransport}. There has been only a few works exploring the Fibonacci model in presence of interactions \cite{varma2019,mace2019,settino}. In Ref.~\cite{mace2019}, the spectral properties of the Fibonacci model were studied as a function of the potential strength at a fixed interaction strength of $\Delta=t_h$. At this interaction strength, in the absence of the Fibonacci potential, i.e., in the ordered XXZ chain, the transport is known to be superdiffusive \cite{stein2017,stein2012,Znidaric2011}. A transition to the MBL phase was predicted. This is very interesting, because, in the absence of interactions, the Fibonacci potential shows no localization. The question, then, is whether this MBL can be seen at lower interaction strengths. The infinite temperature transport properties at small interaction were investigated in Ref.~\cite{varma2019}. This study gave strong evidence that presence of a small interaction makes the transport diffusive at all potential strengths. This again is very nontrivial, because, in absence of interactions there is smooth crossover from ballistic to subdiffusive. This crossover between diffusive to subdiffusive transport also occurs in some disordered interacting systems \cite{Varma_2017}. The question, then, becomes whether transport can become anomalous again at intermediate interaction strengths. One parameter point was shown in favor of this in Ref.~\cite{varma2019}. Our choice of interaction strength, $\Delta=0.5t_h$, is intermediate.   At this choice of interaction strength, in the absence of the Fibonacci potential, i.e, in ordered XXZ chain, the transport is known to be ballistic \cite{stein2017,stein2012,Znidaric2011}. Our findings provide strong evidence that crossover to MBL with increase in the strength of Fibonacci potential can happen even at this interaction strength, and is preceded by a regime of anomalous subdiffusive transport. This answers both the above questions.  This is also very different from a third work, Ref.~\cite{settino}, which studied transport properties of a different model, the spinful Fibonacci model with Fermi-Hubbard interaction, and showed that a localized phase cannot occur in that system, and there will always be slow subdiffusive transport at large interactions and large potential strengths.

More definitive results on the MBL phase would require study of larger systems up to longer times, which is beyond current state-of-the-art numerical techniques. Given the peculiar spectral properties of this class of models, a study of the energy dependence of the transport properties is  a  very promising  direction for  a  subsequent investigation, for example using open systems techniques \cite{energymbl,brenes}. Moreover, all present studies of transport properties are limited to infinite temperature and zero temperature \cite{PhysRevLett.83.3908,PhysRevB.65.014201}. The finite temperature transport properties, as well as the study of thermoelectric behavior \cite{chiar}, are other interesting but challenging  directions for future work.

\section*{Acknowledgements}
We thank M.~T.~Mitchison, M. Brenes, N.~Lo Gullo, and N.~Laflorencie for insightful discussions. The spin configurational basis, the Hamiltonian, and the operators are generated by the open source python package QuSpin\tcite{quspinI}. We acknowledge the provision of computational facilities by the DJEI/DES/SFI/HEA Irish Centre for High-End Computing (ICHEC). This project received funding from the  European Research Council (ERC) under the European Union's Horizon 2020 research and innovation program (Grant Agreement No. 758403). J.G. is supported by a SFI-Royal Society University Research Fellowship. F.P. has received funding from the  European  Union’s  Horizon  2020  research  and  innovation  programme  under  the  Marie Sklodowska-Curie Grant Agreement No. 838773. A.P. is supported by the  European  Union’s  Horizon  2020  research  and  innovation  programme  under  the  Marie Sklodowska-Curie Grant Agreement No. 890884.

\section*{Appendix}
\appendix 
\section{Normalization and restriction to half-filled sector}
\label{app:halfill}
In \teqref{eq:vediapp}, we use the normalized typical state
\begin{equation}
\ket{\tilde{\psi}_{N/2}} = \frac{1}{\sqrt{C}}  \ket{\psi_{N/2}},
\end{equation}
where $C$ is the normalization constant, given by
\begin{align}
C &= \braket{\psi_{N/2}|\psi_{N/2}} = \braket{\psi|\hat{n}_{N/2}|\psi} \nonumber \\ &\approx \overline{\braket{\psi|\hat{n}_{N/2}|\psi}} = \frac{Tr[\hat{n}_{N/2}]}{2^N} = \frac{1}{2}.
\end{align}
Moreover, the subtraction of the factor 1/2 from the term within the parentheses amounts to subtracting the background initial occupation of sites away from the site $N/2$, where the typical state is initially localized. In order to verify it, we notice that for $r \neq N/2$,
\begin{align}
\label{eq:backocc}
n_r(0) &= \braket{\tilde{\psi}_{N/2}|\hat{n}_r |\tilde{\psi}_{N/2}} \approx 2\braket{\psi_{N/2}|\hat{n}_r |\psi_{N/2}} \nonumber \\
&=  2\braket{\psi|\hat{n}_{N/2} \hat{n}_r \hat{n}_{N/2}|\psi} = 2\braket{\psi|\hat{n}_r \hat{n}_{N/2}|\psi} \nonumber \\
&\approx 2\overline{\braket{\psi|\hat{n}_r \hat{n}_{N/2}|\psi}} = 2 \frac{Tr[\hat{n}_r \hat{n}_{N/2}]}{2^N} = \frac{1}{2},
\end{align} 
where in the last equality we use that $Tr[\hat{n}_r \hat{n}_{N/2}]= 2^{N-2}$. 

The above results and those described in Sec.\tref{sec:denscorr} do not make use of the fact that the system is number conserving. For a large enough number conserving system, the biggest contribution to \teqref{eq:Ctyp} comes from the half-filled sector. It is plausible that in such case, one can completely restrict the calculation to the half-filled sector, starting from a typical state in the sector, thus saving computational resources and pushing forward the system size. In complete analogy to Sec.\tref{sec:denscorr}, we define a typical state in the half-filled subsector
\begin{equation}
\ket{\psi^s} = \frac{1}{D^s} \sum\limits_{k=1}^{D^s} c_k \ket{\phi_k^s},  \ \ \ \ \ \ \ D^s = \frac{N!}{(N/2)! (N/2)!},
\end{equation} 
where $\{ \ket{\phi^s_k} \}_{k=1}^{D^s}$ is an orthonormal basis in the half-filled sector. The new normalization constant $C^s$ in
\begin{equation}
\ket{ \tilde{\psi}^s_{N/2}} = \frac{1}{\sqrt{C^s}} \ket{\psi^s_{N/2}}, \ \ \ \ \ \ \ \ket{\psi^s_{N/2}} = \hat{n}_{N/2} \ket{\psi^s},
\end{equation}
is given by
\begin{align}
C^s &= \braket{\psi^s_{N/2}|\psi^s_{N/2}} = \braket{\psi^s| \hat{n}_{N/2}|\psi^s} \approx \overline{\braket{\psi^s| \hat{n}_{N/2}|\psi^s}} \nonumber \\
&=\frac{Tr[\hat{n}_{N/2}]}{D^s} = \frac{(N-1)!}{ \frac{N-1}{2}!\frac{N-1}{2}!} \frac{\frac{N}{2}!\frac{N}{2}!}{N!} = \frac{1}{2}. 
\end{align}
As before, we have
\begin{equation}
\ket{\tilde{\psi}^s_{N/2}} \approx \sqrt{2} \ket{\psi_{N/2}^s}.
\end{equation}
However, the background occupation of sites $q \neq N/2$ is now less than 1/2, as it is possible to notice by reproducing the result of \teqref{eq:backocc} in the half-filled sector:
\begin{align}
n^s_r(0) &= \braket{\tilde{\psi}^s_{N/2}|\hat{n}_r |\tilde{\psi}^s_{N/2}} \approx 2\braket{\psi^s_{N/2}|\hat{n}_r |\psi^s_{N/2}} \nonumber \\
&= 2\braket{\psi^s|\hat{n}_r \hat{n}_{N/2}|\psi^s} \nonumber \approx 2\overline{\braket{\psi^s|\hat{n}_r \hat{n}_{N/2}|\psi^s}} \nonumber \\ &= 2 \frac{Tr[\hat{n}_r \hat{n}_{N/2}]}{D^s} =
 \frac{(N-2)!}{ \frac{N-2}{2}!\frac{N}{2}!} \frac{\frac{N}{2}!\frac{N}{2}!}{N!}   
 \nonumber \\ &=\frac{1}{2} \Bigl( 1 - \frac{1}{N-1} \Bigr).
\end{align} 
Finally, in analogy with \teqref{eq:vediapp} we are able to define
\begin{align}
C^s_l(t) &= \frac{1}{2}( n^s_l(t) - n^s_l(0)), \ \ \ \ l \neq N/2 \nonumber \\
& \approx \frac{1}{2} \Bigl[ n^s_l(t) -\frac{1}{2} \Bigl( 1 - \frac{1}{N-1} \Bigr) \Bigr],
\end{align}
where $n^s_l(t)$ is the expectation value of the operator $\hat{n}_{l}$ at time $t$, starting from the initial state given by $\ket{\tilde{\psi}^s_{N/2}}$. For a large enough system, we expect $C^s_l(t) \approx C_l(t)$. By directly comparing simulations performed for a short time interval on a chain of size $N=20$ both in the total Hilbert space and in the largest sector at half-filling, we have confirmed our conjecture.

\section{Relation with Green-Kubo conductivity}
\label{app:Green-Kubo}

The Green-Kubo formula for particle conductivity at finite temperature can be written as
\begin{align}
\label{Green_Kubo}
\sigma_{GK} = \beta\lim_{t\rightarrow\infty}\lim_{N\rightarrow\infty} \frac{1}{N} {\rm Re}\left(\int_{0}^{t} dt^\prime \langle \hat{I}(t^\prime) \hat{I} \rangle_{\beta,\mu}\right),
\end{align}
where $\hat{I}$ is the particle current operator and $\langle ... \rangle_{\beta,\mu}$ denotes the average taken over the thermal state of the system with temperature $\beta$ and chemical potential $\mu$. In the above the order of limits cannot be interchanged. For one-dimensional systems with open boundary condition, the particle current operator is given by
\begin{align}
\hat{I}=\frac{d\hat{x}}{dt},
\end{align}
where
\begin{align}
\hat{x}=\sum_{p=1}^N p\hat{n}_p
\end{align}
is the position operator. This definition gives,
\begin{align}
\langle \hat{I}(t_1) \hat{I}(t_2) \rangle=\frac{d}{dt_1}\frac{d}{dt_2}\left(\sum_{p,q=1}^N pq \langle \hat{n}_p(t_1)\hat{n}_q(t_2)\rangle_{\beta,\mu} \right).
\end{align} 
Using time translational invariance of the thermal state and changing the variable to $\tau=t_1-t_2$, we have
\begin{align}
\langle \hat{I}(\tau) \hat{I} \rangle=-\frac{d^2}{d\tau^2}\left(\sum_{p,q=1}^N pq C_{pq}(\beta,\mu,t) \right),
\end{align}
where
\begin{align}
 C_{pq}(\beta,\mu,\tau)=\langle \hat{n}_p(\tau)\hat{n}_q\rangle_{\beta,\mu}.
\end{align}
Now we use the relation $2pq=p^2+q^2-(p-q)^2$, along with the assumption that the Hamiltonian is number conserving, so that $\frac{d}{d\tau}\left(\sum_{p=1}^N \hat{n}_p(\tau)\right)=0$, to obtain
\begin{align}
&\langle \hat{I}(\tau) \hat{I} \rangle_{\beta,\mu}=\frac{1}{2}\frac{d^2}{d\tau^2}\left(\sum_{p,q=1}^N (p-q)^2 C_{pq}(\beta,\mu,\tau) \right). \nonumber \\
& \int_{0}^{\tau} dt^\prime \langle \hat{I}(t^\prime) \hat{I} \rangle_{\beta,\mu} = \frac{1}{2}\frac{d}{d\tau}\left(\sum_{p,q=1}^N (p-q)^2 \langle C_{pq}(\beta,\mu,\tau) \right).
\end{align}
Using the above equation in Eq.(\ref{Green_Kubo}), we have
\begin{align}
&\sigma_{GK} = \frac{\beta}{2}\lim_{t\rightarrow\infty} \frac{d}{dt} m_2^{nn}(t)\nonumber \\
\label{def_m2nn}
&m_2^{nn}(t)=\lim_{N\rightarrow\infty} \frac{1}{N} {\rm Re}\left(\sum_{p,q=1}^N (p-q)^2 C_{pq}(\beta,\mu,t) \right).
\end{align}
Further simplification of $m_2^{nn}(t)$ is possible if the system has translational invariance in the thermodynamic limit. In that case, $C_{pq}(t)$ becomes almost independent of $q$ for large enough system sizes. So, we can fix $q=N/2$, to obtain,
\begin{align}
m_2^{nn}(t)=\lim_{N\rightarrow\infty} {\rm Re}\left(\sum_{p=1}^{N} \left(p-\frac{N}{2}\right)^2 C_{p\frac{N}{2}}(\beta,\mu,t) \right),
\end{align}
Writing the above expression in the high-temperature limit, $\beta\rightarrow 0$, using the dynamical typicality and the fact that at $\beta\rightarrow 0$, $m_2^{nn}(t)$ is real, we have, in the presence of translational invariance in the thermodynamic limit, 
\begin{align}
\sigma^2(t) = 4\lim_{\beta \rightarrow 0} m_2^{nn}(t).
\end{align}
The scaling of the quantity $\sigma^2(t)$ with time gives the nature of high temperature transport. Let us define the time-dependent diffusion coefficient at high temperature as
\begin{align}
\mathcal{D}(t)=\lim_{\beta \rightarrow 0} \lim_{N\rightarrow \infty}\frac{1}{N}\int_{0}^{t} dt^\prime \langle \hat{I}(t^\prime) \hat{I} \rangle_{\beta,\mu}.
\end{align}
Then, from above,
\begin{align}
\frac{d\sigma^2}{dt}=8\mathcal{D}(t).
\end{align}
This derivation relies on translational invariance of the system in the thermodynamic limit. Though the Fibonacci model is not translationally invariant in the thermodynamic limit, the translational invariance is effectively restored on averaging over the various realizations.

\section{Relation with the spread of wavepacket in the noninteracting system}
\label{app:relation_with_wavepacket}

We go back again to the case of the noninteracting system, described by the Hamiltonian in \teqref{eq:hni}, that we diagonalize as
\begin{equation}
\bm{\Phi}^T\bm{H}_{NI}\bm{\Phi} = \bm{D}, \ \ \ \ \ \ \ \bm{D} = \text{diag}\{ \epsilon_{\nu} \},
\end{equation}
with the single-particle eigenvectors given by the columns of $\bm{\Phi}$ and the eigenvalues by $\{\epsilon_{\nu}\}$. In the diagonalized basis, the Hamiltonian reads
\begin{equation}
\hat{H}_{\text{NI}} = \sum\limits_{\nu=1}^{N} \epsilon_{\nu} \hat{c}_{\nu}^{\dagger} \hat{c}_{\nu},
\end{equation}
where $\hat{c}_{\nu} = \sum_{p=1}^N \Phi_{p\nu} \hat{a}_p$ are the fermionic annihilation operators in the eigenbasis. The two time density correlation at finite temperature can be simplified as follows
\begin{align}
\label{eq:corrtemp}
&C_{pq}(\beta,t) = \braket{\hat{n}_p(t)\hat{n}_q} - \braket{\hat{n}_p}\braket{\hat{n}_q} \nonumber \\
&= \sum\limits_{\nu,\alpha=1}^N \Phi_{\nu p} \Phi_{\nu q} \Phi_{\alpha p} \Phi_{\alpha q} e^{i t/\hbar(\epsilon_{\nu}-\epsilon_{\alpha})} [1-n_F(\epsilon_{\alpha})] n_F(\epsilon_{\nu}),
\end{align}
where $\braket{}$ indicates the ensemble average, after applying the Wick's theorem
\begin{align}
&\braket{ \hat{a}_p^{\dagger}(t_p)\hat{a}_q(t_q) \hat{a}_m^{\dagger}(t_m)  \hat{a}_n(t_n)} \nonumber \\
 &= \braket{ \hat{a}_p^{\dagger}(t_p) \hat{a}_q(t_q) } \braket  {\hat{a}_m^{\dagger}(t_m)  \hat{a}_n(t_n)} \nonumber \\
 &+  \braket{ \hat{a}_p^{\dagger}(t_p) \hat{a}_n(t_n) } \braket  {\hat{a}_q(t_q)  \hat{a}^{\dagger}_m(t_m)},
\end{align}
and the following relations
\begin{align}
 \braket{ \hat{a}_p^{\dagger}(t_p) \hat{a}_q(t_q) } &= \sum\limits_{\nu=1}^N \Phi_{\nu p} \Phi_{\nu q} e^{i \epsilon_{\nu}/\hbar(t_p-t_q)} n_F(\epsilon_{\nu}) \nonumber \\
  \braket{ \hat{a}_p(t_p) \hat{a}_q^{\dagger}(t_q) } &= \sum\limits_{\nu=1}^N \Phi_{\nu p} \Phi_{\nu q} e^{i \epsilon_{\nu}/\hbar(t_p-t_q)}(1- n_{F}(\epsilon_{\nu}))
\end{align}
with $n_F(E)=\{1 + \exp[\beta(E - \mu)]\}^{-1}$ the Fermi-Dirac distribution.
Now, we take the infinite temperature limit $\beta \rightarrow 0$ of \teqref{eq:corrtemp}, and shift the labels to consider the correlation between the middle of the chain and the other sites $q = N/2$ and $p = l+N/2$ as in the previous section,
\begin{align}
\label{eq:equival}
C_l(t) &= \frac{1}{4} \sum\limits_{\nu,\alpha=1}^N \Phi_{\nu l} \Phi_{\nu N/2} \Phi_{\alpha l} \Phi_{\alpha N/2} e^{i t(\epsilon_{\nu}-\epsilon_{\alpha})/\hbar} \nonumber \\
&\frac{1}{4} \Bigl( \sum\limits_{\alpha=1}^N  \Phi_{\alpha l} \Phi_{\alpha N/2} e^{-it\epsilon_{\alpha}/\hbar} \Bigr)  \Bigl( \sum\limits_{\nu=1}^N \Phi_{\nu l} \Phi_{\nu N/2} e^{it\epsilon_{\nu}/\hbar} \Bigr) \nonumber \\
&= \frac{1}{4} |\Psi_l(t)|^2,
\end{align}
where
\begin{equation}
\Psi_l(t) = \sum\limits_{\nu=1}^N  \Phi_{\nu l} \Phi_{\nu N/2} e^{it\epsilon_{\nu}/\hbar}.
\end{equation}
The dynamics of each $\Psi_l(t)$ corresponds to evolution according to
\begin{equation}
i\frac{d \Psi_l(t) }{dt} = \sum\limits_{r=1}^N \bm{H}_{lr} \Psi_l(t),
\end{equation} 
starting from the initial condition $\Psi_l(t) = \delta_{l N/2}$. Thus, from \teqref{eq:equival} we derive that the two time density correlation at infinite temperature we use to classify transport in the interacting system is directly proportional to $|\Psi_l(t)|^2$ in single-particle systems:
\begin{equation}
|\Psi_l(t)|^2 = 4C_l(t) \approx 4C^{typ}_l(t).
\end{equation} 
Physically, $|\Psi_l(t)|^2$ gives the probability of finding a particle at site $l=p-N/2$, after initializing the system with a single particle located at site $N/2$. From above, and Eqs.(\ref{eq:dx2}) and (\ref{eq:sigsq}), we see that, for a noninteracting system, $\sigma^2(t)=\Delta x^2(t)$. But unlike $\Delta x^2(t)$, $\sigma^2(t)$ is well defined also in the presence of interactions.

\bibliographystyle{apsrev4-1}
\bibliography{fib}

\end{document}